\documentclass[aps,prb,twocolumn,groupedaddress,amssymb,superscriptaddress]{revtex4-1}
\usepackage{graphicx}
\usepackage{epstopdf}
\usepackage{dcolumn}
\usepackage{bm}
\usepackage{color}
\usepackage{amsmath}
\usepackage{charter}
\usepackage{amssymb}
\usepackage{setspace}
\usepackage{amsmath}
\usepackage{csquotes}
\usepackage{float}
\usepackage {braket}
\usepackage{hyperref}
\usepackage{xcolor}
\usepackage{xfrac}
\usepackage{siunitx}
\usepackage{tabularx}
\usepackage{tabulary}
\usepackage{tabu}
\usepackage{mathtools}
\usepackage{xspace}
\usepackage{booktabs}
\usepackage{titlesec}
\usepackage{multirow}
\begin{document}
\title[Magnetically driven loss of centrosymmetry in metallic Pb$_2$CoOsO$_6$]{Magnetically driven loss of centrosymmetry in metallic Pb$_2$CoOsO$_6$}

\author{A. J. Princep}
\email[]{princep@physics.ox.ac.uk}
\affiliation{Department of Physics, University of Oxford, Clarendon Laboratory, Parks Road, Oxford, OX1 3PU, United Kingdom}
\affiliation{ISIS Facility, Rutherford Appleton Laboratory, Chilton, Didcot, OX11 0QX, United Kingdom}
\author{H. L. Feng}
\email[]{hai.feng@iphy.ac.cn}
\affiliation{National Institute for Materials Science, 1-1 Namiki, Tsukuba, Ibaraki 305-0044, Japan}
\affiliation{Beijing National Laboratory for Condensed Matter Physics, Institute of Physics, Chinese Academy of Sciences, Beijing 100190, China}
\author{Y. F. Guo}
\email[]{guoyf@shanghaitech.edu.cn}
\affiliation{School of Physical Science and Technology, ShanghaiTech University, Shanghai 201210, China}
\author{F. Lang}
\affiliation{Department of Physics, University of Oxford, Clarendon Laboratory, Parks Road, Oxford, OX1 3PU, United Kingdom}
\author{H. M. Weng}
\affiliation{Beijing National Laboratory for Condensed Matter Physics, Institute of Physics, Chinese Academy of Sciences, Beijing 100190, China}
\affiliation{Collaborative Innovation Center of Quantum Matter, Beijing, China }
\author{P. Manuel}
\author{D. Khalyavin}
\affiliation{ISIS Facility, Rutherford Appleton Laboratory, Chilton, Didcot, OX11 0QX, United Kingdom}
\author{A. Senyshyn}
\affiliation{FRM-II, Technische Universität München, Garching 85747, Germany}
\author{M. C. Rahn}
\affiliation{Department of Physics, University of Oxford, Clarendon Laboratory, Parks Road, Oxford, OX1 3PU, United Kingdom}
\affiliation{Institute for Solid State and Materials Physics, Technical University of Dresden, 01062 Dresden,Germany}
\author{Y. H. Yuan}
\author{Y. Matsushita}
\affiliation{National Institute for Materials Science, 1-1 Namiki, Tsukuba, Ibaraki 305-0044, Japan}
\author{S. J. Blundell}
\affiliation{Department of Physics, University of Oxford, Clarendon Laboratory, Parks Road, Oxford, OX1 3PU, United Kingdom}
\author{K. Yamaura}
\email[]{Yamaura.kazunari@nims.go.jp}
\affiliation{National Institute for Materials Science, 1-1 Namiki, Tsukuba, Ibaraki 305-0044, Japan}
\affiliation{Graduate School of Chemical Sciences and Engineering, Hokkaido University, Sapporo, Hokkaido 060-0810, Japan}
\author{A. T. Boothroyd}
\email[]{a.boothroyd@physics.ox.ac.uk}
\affiliation{Department of Physics, University of Oxford, Clarendon Laboratory, Parks Road, Oxford, OX1 3PU, United Kingdom}

\date{\today}

\begin{abstract}

We report magnetic, transport, neutron diffraction, and muon spin rotation data showing that Pb$_2$CoOsO$_6$, a newly synthesized metallic double-perovskite with a centrosymmetric space group at room temperature, exhibits a continuous second-order phase transition at 45 K to a magnetically ordered state with a non-centrosymmetric space group. The absence of inversion symmetry is very uncommon in metals, particularly metallic oxides. In contrast to the recently reported ferroelectric-like structural transition in LiOsO$_3$, the phase transition in Pb$_2$CoOsO$_6$ is driven by a long-range collinear antiferromagnetic order, with propagation vector $\textbf{k} = (\frac{1}{2},0,\frac{1}{2})$, which relieves the frustration associated with the symmetry of themagnetic exchanges. This magnetically-driven loss of inversion symmetry represents a new frontier in the search for novel metallic behavior.  

\end{abstract}




\maketitle

\section{Introduction}
Metals whose crystal structure lacks a center of inversion symmetry have been attracting increasing interest owing to the novel phenomena they can exhibit, such as optical activity \cite{mineev_opticalactivity} and a highly anisotropic thermopower, (a desirable property of certain thermoelectric devices) \cite{ncomms_puggioni}. Non-centrosymmetric metals having strong electronic correlations can support exotic emergent quasiparticles, including skyrmions in chiral magnets \cite{muhlbauer_science}. Non-centrosymmetric superconductors are of particular interest because they can have spin-polarized supercurrents and unconventional pairing states with mixed singlet-triplet character even in the absence of strong electronic correlations \cite{fujimoto_jpsj,NCSM_book, samokhin_annals}. The key feature of non-centrosymmetric systems is a band splitting throughout much of momentum space caused by spin–orbit coupling (Dresselhaus splitting \cite{dresselhaus_prl}) which leads to a non-trivial topology of the electronic wave functions and plays an essential role in all these phenomena.
  
Non-centrosymmetric metals (NCSM) are relatively uncommon. This is because conduction electrons can effectively screen the electric dipole formation which is generally associated with acentricity. In fact, there exist to date only around 30 known NCSM, of which only a handful are metallic oxides \cite{ncomms_puggioni}. Nevertheless, oxides are particularly attractive for device applications due to their stability under typical operating conditions, so the discovery of more oxide NCSM would be of considerable interest. 

Structural transitions in metallic oxides which remove the center of inversion symmetry have previously been observed in Cd$_2$Re$_2$O$_7$ \cite{sergienko_prl} and LiOsO$_3$ \cite{shi_nmater}. The origin and nature of the transition in Cd$_2$Re2O$_7$ is uncertain, but LiOsO$_3$ was found to be an example of what Anderson and Blount referred to as a “ferroelectric metal”, i.e. a metal having a continuous structural phase transition accompanied by the appearance of a polar axis and the disappearance of an inversion centre \cite{andersonblount}. The phase transition in LiOsO$_3$ is driven by the ordering of Li ion displacements \cite{shi_nmater}. 

Here, we report structural, magnetic and transport measurements of Pb$_2$CoOsO$_6$, another metal that undergoes a phase transition to a non-centrosymmetric structure. In this case, however, we show that the loss of inversion symmetry is driven by magnetoelastic coupling to a pair of antiferromagnetic order parameters which relieve the magnetic frustration of the higher symmetry phase, analogous to the behavior of a type-II hybrid-improper multiferroic \cite{hybridimproper1, hybridimproper2}. This represents a new paradigm for obtaining NCSM using magnetic frustration as a key ingredient.

\section{Structural and bulk characterisation}
Polycrystalline and single crystal Pb$_2$CoOsO$_6$ was prepared by a high pressure method (see Supplementary Materials). The room temperature crystal structure of polycrystalline Pb$_2$CoOsO$_6$ was initially determined by synchrotron X-ray diffraction (XRD) (see Supplementary Materials). The structure found is the fully ordered double perovskite structure with space group of P2$_1$/$n$ (tilt pattern a$^-$a$^-$c$^+$ in the notation of Glazer \cite{glazer_notation}). The Co and Os atoms fully occupy the Wyckoff positions 2a and 2b, respectively, with no site mixing observed to the accuracy of the measurement. The bond valence sums imply +2 and +6 valence states for the Co and Os atoms respectively \cite{ederer_prb, azuma_jacs}. The refined crystal structure is depicted in Fig. 1(a). The degree of distortion from the cubic structure is indicated by the deviation of inter-octahedral Co-O-Os bond angles which are 180$^\circ$ in the ideal structure, but here we find them to be 168$^\circ$, 172$^\circ$, and 145$^\circ$, respectively, implying substantial buckling of the octahedral connections. 

The temperature variation of the structure of Pb$_2$CoOsO$_6$ was investigated between 1.5 K and 300 K by neutron powder diffraction (NPD). Neutron diffraction experiments were performed on a 4g powder at the ISIS facility on the WISH diffractometer. Rietveld refinements were carried out using the FullProf suite \cite{fullprof} using the magnetic form factor for Os$^{6+}$ determined by Kobayashi \emph{et. al.} \cite{5d_formfactors}. Results of refinement against the NPD pattern at 300 K are fully consistent with the synchrotron XRD results, including the absence of any Co/Os site mixing to within the experimental uncertainty of $\approx$1\%. The lattice parameters decrease monotonically with temperature below 300 K until $T_{\rm N}$ = 45 K, below which the parameters $a$ and $b$ increase slightly while $c$ decreases significantly for a small net reduction in the unit-cell volume [see Fig. 1(b)]. This abrupt change in behavior of the lattice parameters is typically indicative of a magnetoelastic structural distortion accompanying the magnetic order, as discussed below. We identify $T_{\rm N}$ with a bulk antiferromagnetic ordering transition, based on the neutron diffraction and muon spin rotation experiments presented below, as well as by the existence of characteristic signatures in other physical properties (see Fig. 2).

Fig. 2(a) presents the temperature dependence of the electrical resistivity $\rho$ of Pb$_2$CoOsO$_6$ measured on a single crystal between 2 K and 300 K. The single crystal was confirmed to have the same crystal structure as the powder sample by single crystal XRD (see Supplementary Materials and Fig. S2). Above $T_{\rm N}$, $\rho$ is roughly linear in $T$, and displays a pronounced change at T$_{\rm N}$ without fundamentally altering its metallic character.  The drop in $\rho$ below $T_{\rm N}$ is likely due to the reduction in scattering of the conduction electrons by paramagnetic fluctuations as the system undergoes magnetic ordering. At the lowest temperatures there is a small upturn in the resistivity, however given the resistivity decreases monotonically over several decades in temperature and there is no evidence for a further transition we attribute this to a simple experimental artefact arising from the contacts with the sample. The dc magnetic susceptibility ($\chi$ vs. $T$) of Pb$_2$CoOsO$_6$, presented in Fig. 2(b), exhibits a sharp peak at $T_{\rm N}$, implying that the transition has a magnetic origin. A Curie-Weiss fit to $\chi^{-1}$ in the paramagnetic region above 200 K (see Fig. S2a in Supplementary Materials) yields the Weiss temperature $\Theta_{\rm W} \approx$ $-$106 K, suggesting dominant antiferromagnetic (AFM) interactions and a  $|\Theta_{\rm W} $/ $T_{\rm N}| \approx 2.2$, indicative of weak magnetic frustration. The effective magnetic moment ($\mu_{\rm eff}$) per formula unit is approximately 4.9 $\mu_{\rm B}$.

\begin{figure}
\includegraphics[width=0.5\textwidth]{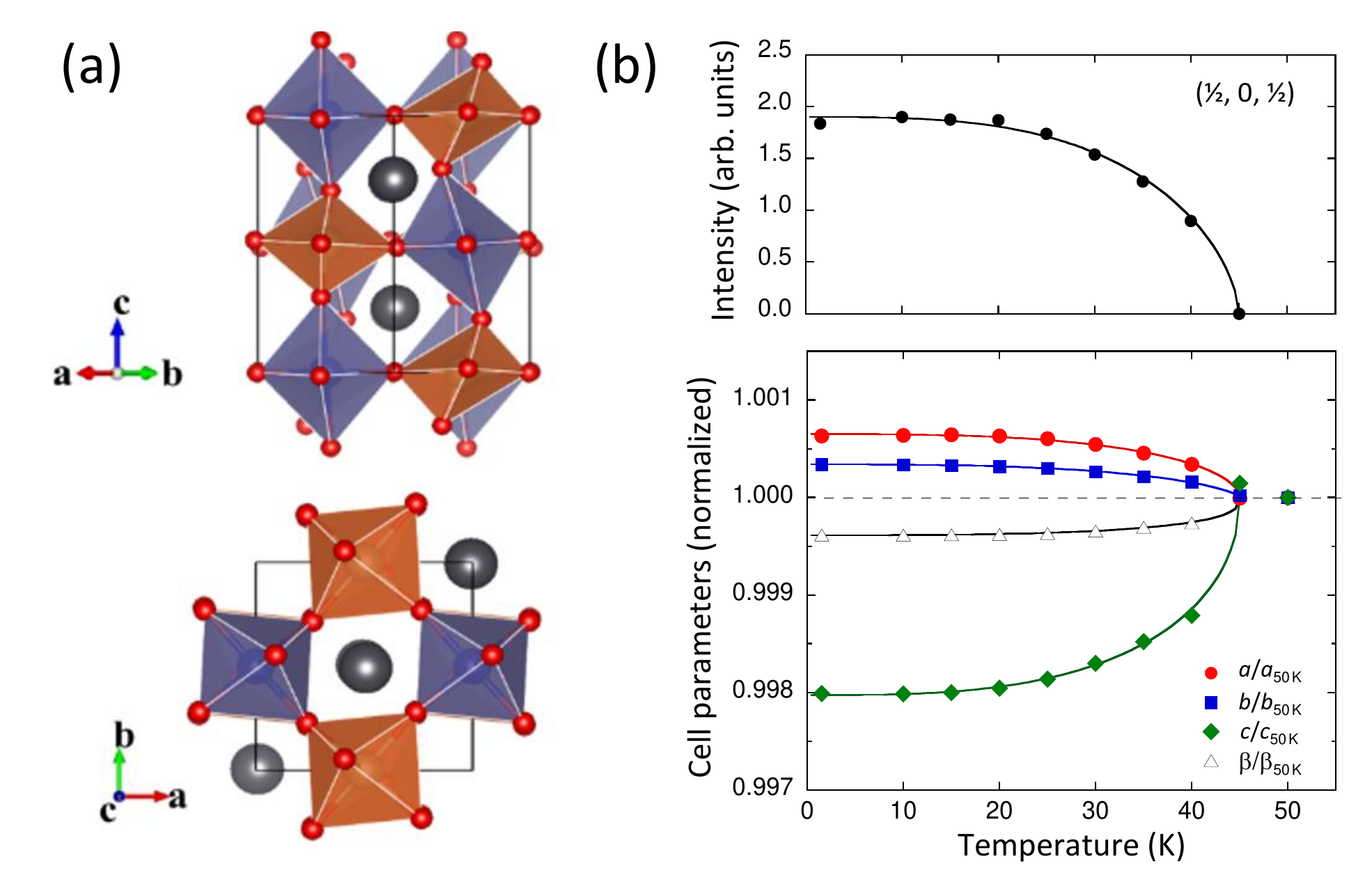}
\caption{\label{Fig1}  (a) Crystallographic views of Pb$_2$CoOsO$_6$ along the [110] (left) and [001] (right) directions. The blue and brown octahedra represent CoO$_6$ and OsO$_6$, respectively. The large grey solid spheres represent Pb. (b) (upper) temperature dependence of the fundamental magnetic reflection measured in neutron scattering and (lower) temperature dependence of the lattice parameters below the ordering temperature, relative to their value at $T_{\rm N}$. Solid lines are a guide to the eye }
\end{figure}

\begin{figure}
\includegraphics[width=0.45\textwidth]{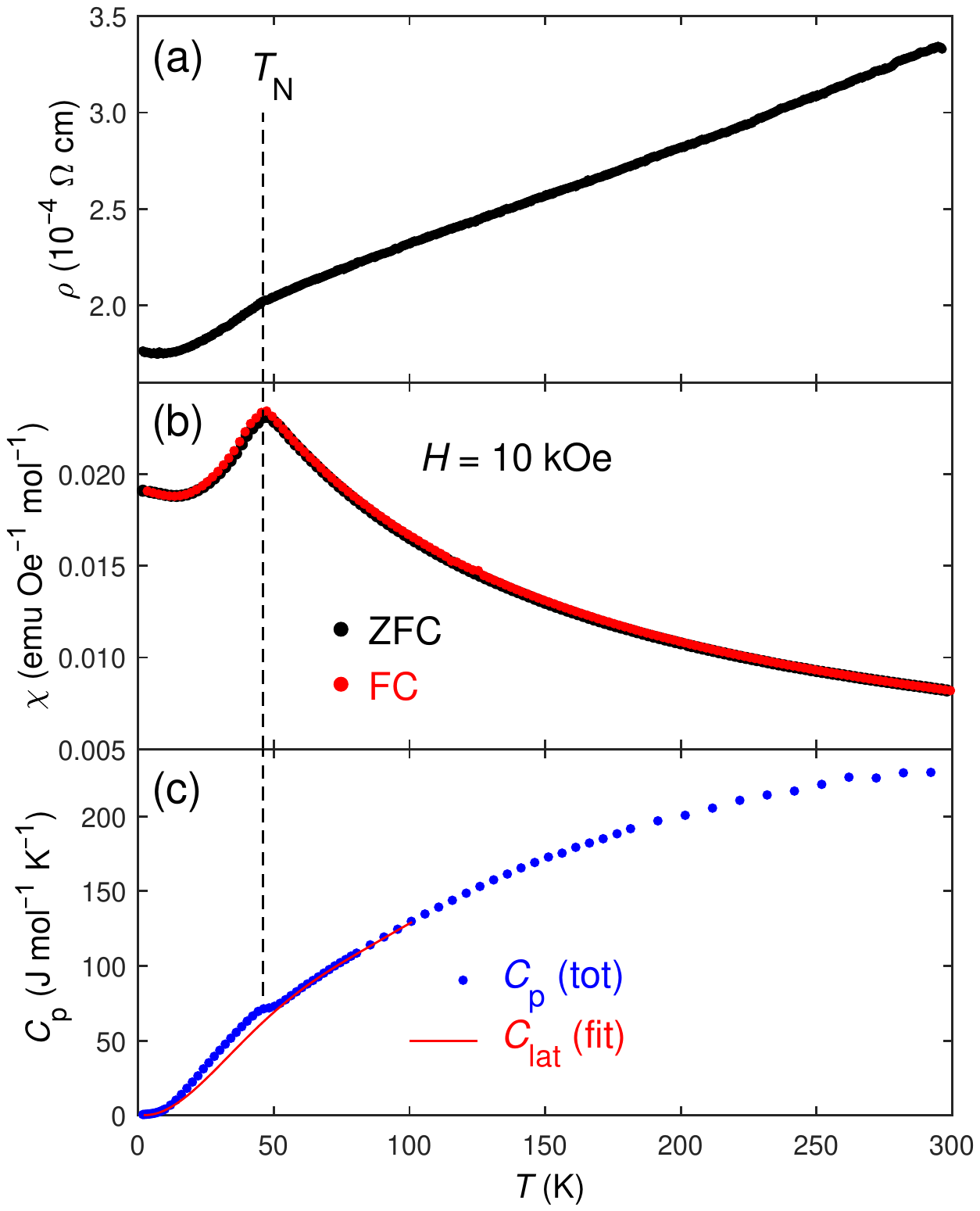}
\caption{\label{Fig2} Temperature dependence of bulk properties of Pb$_2$CoOsO$_6$. (a) Resistivity $\rho$ of a single crystal. (b) Magnetic susceptibility $\chi$ of the polycrystalline sample. (c) Heat capacity $C_{\rm p}$ of the polycrystalline sample.}
\end{figure}

\begin{figure}
\includegraphics[width=0.5\textwidth]{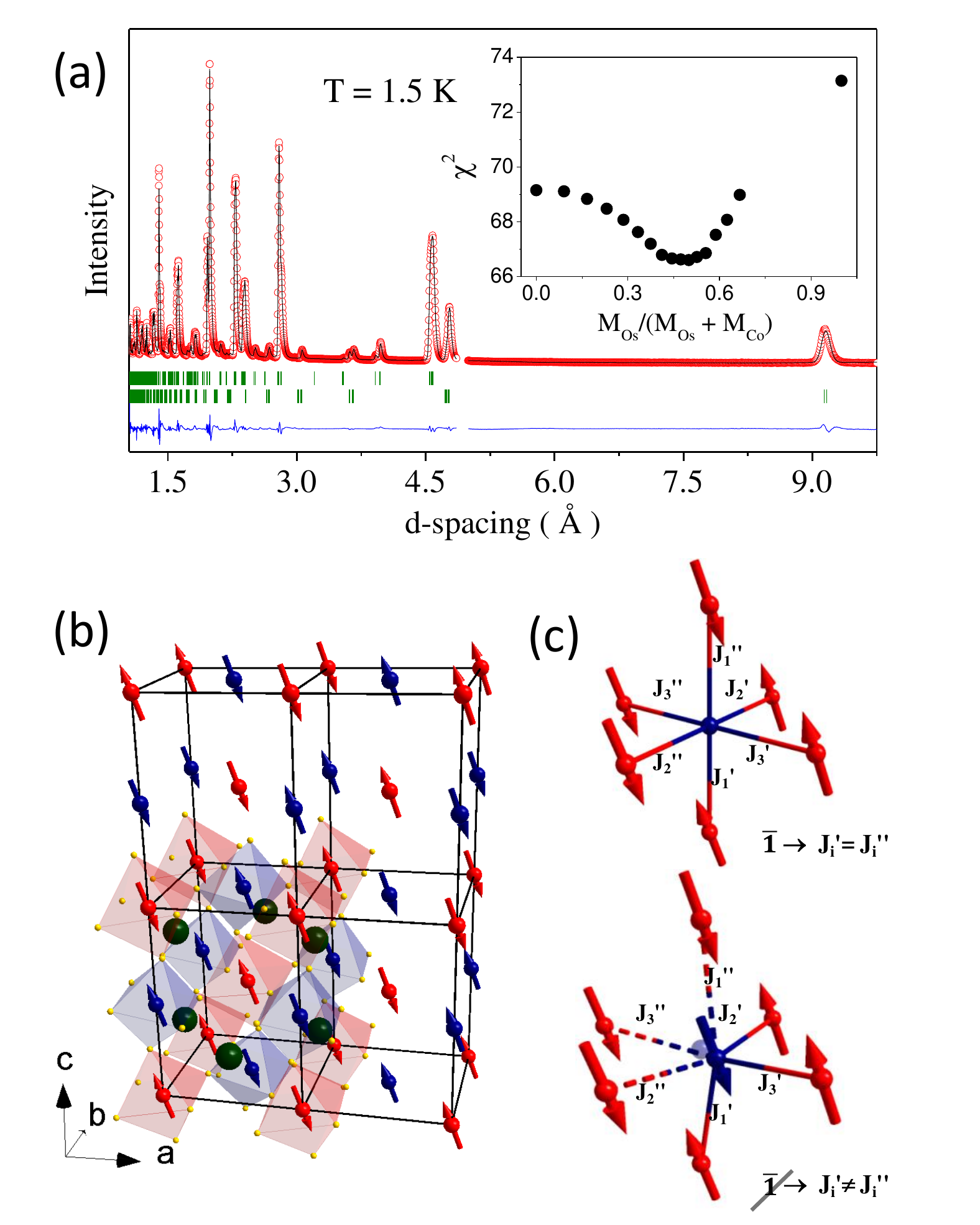}
\caption{\label{Fig3} Magnetic structure of Pb$_2$CoOsO$_6$. (a) Neutron diffraction data recorded in banks 2 and 9 of the WISH diffractometer are shown together with the Rietveld fit. The data are in red, the fit in black, and the difference in blue. The green tick marks indicate structural (upper) and magnetic (lower) Bragg peaks. The inset shows how the goodness-of-fit statistic $\chi^2$ for the Rietveld fit varies as a function of the fraction of the total ordered moment that is located on the Os site. (b) Refined magnetic structure ($T$ $<$ $T_{\rm N}$), with red and blue arrows as the Co$^{2+}$ and Os$^{6+}$ moments, respectively. The red and blue octahedra represent the oxygen coordination polyhedra, and the dark green spheres depict non-magnetic Pb$^{2+}$. (c) Absent a structural distortion, the mean field exerted by nearest neighbour atoms cancels by symmetry. Removing the center of inversion symmetry by displacing the Co and Os sublattices relative to one-another permits a non-zero nearest neighbour coupling. }
\end{figure}

\section{Magnetic structure refinement}
Below $T_{\rm N}$, additional peaks were observed in NPD patterns that could be indexed with a propagation vector $\textbf{k} =(\frac{1}{2},0,\frac{1}{2})$ [see Fig. 1(b) upper and Fig. 3(a)]. There are four magnetic irreducible representations (irreps) that are compatible with $k =(\frac{1}{2},0,\frac{1}{2})$ in the space group P2$_1$/$n$ and these are the 1-dimensional irreps mY$_1^+$, mY$_2^+$, mY$_1^-$, and mY$_2^-$ \cite{miller_love}. The ``+'' type irreps correspond to order on only the Co sites, and the ``–'' irreps to order on only the Os sites, and each irrep has a specific relation between the direction of the moments on the two equivalent metal sites. The allowed magnetic Bragg reflections vary with the choice of irrep, and inspection of the NPD data allowed us to constrain the possible irreps to mY$_1^+$ and mY$_2^-$. A model combining these two irreps was refined against NPD data from banks 2-9 of WISH. Fig. 3(a) shows data from banks 2 and 9 recorded at 1.5 K together with the refinement. The results of the refinement are tabulated in Table 1, and the magnetic structure is shown in Fig. 3(b). A direct observation of the loss of centrosymmetry by measuring of optical second harmonic generation or Friedel pairs would be unlikely to be fruitful, as the size of the distortion is likely to be as large or smaller than that in magnetically induced (type-II) multiferroics where a similar mechanism pertains. There, the size of the polar distortion is of order 10fm (i.e. the atomic nucleus) and thus far too small to detect by standard methods \cite{walker2011}. 

The parameters describing the Co and Os moments are strongly correlated in the refinement and constraints are needed to reach convergence. This is specific to $\textbf{k} =(\frac{1}{2},0,\frac{1}{2})$ in the space group P2$_1$/$n$ where both magnetic atoms contribute intensity to the reflections in the exact same way except for their different form factors. Initial refinements performed with moments on only the Co sites or only the Os sites indicated that the magnetic structure is collinear and that the component along the $b$-axis ($M_b$) is undetectably small. Thereafter, we set $M_b$ to zero and performed fits as a function of the ratio of the moments on the Co and Os sites assuming all moments to be collinear. The substantial difference between the magnetic form factors of Co$^{2+}$ and Os$^{6+}$ provides a degree of sensitivity to the magnetic moment on each sublattice, and the best fit was found with approximately the same moment (2.04 $\mu_{\rm B}$) on both Co and Os sites inclined at an angle of about 22 deg to the c axis. Fig. 3(a) shows data from banks 2 and 9 recorded at 1.5 K together with the refinement, and the insert shows the fit quality as the Os moment fraction is varied. The results of the refinement are tabulated in Table 1, and the magnetic structure is shown in Fig. 3(b). 

The observation of a single magnetic transition indicates coincident order of the two sub-lattices, as observed recently in the osmate double-perovskite Sr$_2$FeOsO$_6$ \cite{kumarpaul}. Coincident order of coupled sublattices is to be expected unless the Co-O-Os superexchange was either highly frustrated or suppressed, in which case one expects two distinct magnetic transitions as appears to be the case in Sr$_2$CoOsO$_6$ \cite{morrow_jacs, magnetoelasticosmate}. It is worthing that the conclusions of the preceding symmetry analysis are not contingent on a single magnetic phase transition. If in reality one of the magnetic sublattices orders at some temperature T$_{\rm N'}$ just below $T_{\rm N}$ the resulting state would still be without a centre of symmetry and thus a magnetically driven polar metal. 

As was noted earlier, below $T_{\rm N}$ the temperature dependence of the lattice parameters abruptly changes and they display an order-parameter-like behavior, suggesting that the lattice distortion is coupled to the magnetic order parameter. The magnetic structure refinement and electronic structure calculations (see below) provide evidence that there are ordered moments on both the Co and Os sites, and moreover, the refined magnetic structure does not possess a centre of symmetry. Any non-zero magneto-elastic coupling will then remove the centre of symmetry of the crystal structure, Fig. 3(c), relieving frustration and reducing the space group symmetry of the structure to the (non-centrosymmetric) polar group P$n$ \cite{isodistort}, with the corresponding Shubnikov group for the magnetic structure being P$_a$c. Both Co and Os have a non-zero orbital angular momentum and so such coupling is expected to be non-negligible, and is indicated by the temperature-dependent variation of the lattice parameters described earlier [see Fig. 1(a)]. This scenario is opposite to the case where relieving orbital degeneracy results in a structural phase transition (i.e. a Jahn-Teller transition) for which any eventual magnetic ordering is expected to occur at a lower temperature  \cite{jahnteller}. Instead, the onset of magnetic order, via spin-orbit coupling, enforces some preferred orbital occupation which then results in a structural distortion \cite{jahnteller} as is the case, for example, at $T_{\rm{N}1}$ in Sr$_2$CoOsO$_6$ \cite{morrow_jacs, magnetoelasticosmate}. Although clear changes in structural elements of the diffraction pattern were observed on cooling through $T_{\rm N}$, it proved impossible to refine a detailed model for the low temperature crystal structure owing to instabilities that resulted from the large number of free parameters.

\begin{table}
\caption{Refined values of the crystal and magnetic structures at 1.5K from neutron powder diffraction. Space group P2$_1$/$n$ ( \# 14, origin choice 2) $a$ = 5.63651(1) \AA, $b$ = 5.58361(7) \AA, $c$ = 7.82321(1) \AA, and $\beta$ = 89.815(2)$^\circ$. Magnetic moments on Co and Os sites were constrained to be equal. Final $R$ values are: $R_{\rm nuc}$ = 2.29\% and $R_{\rm mag}$ = 7.91\%. The lattice vectors of the Shubnikov group in the standard setting P$_ac$ are related to the high temperature crystallographic space group P$2_1/n$ as $(-2,0,0)$, $(0,-1,0)$, $(1,0,1)$ with an origin shift of $(0,\frac{1}{4},0)$.}
\begin{tabular}{ c c c c c c } 
 \hline
Site & Wyck & $x$ & $y$ & $z$ & $B_{\rm iso}$ (\AA$^2$) \\ 
\hline
Pb & 4e & 0.0059(3) & 0.5127(4) & 0.2504(3) & 0.45(4) \\
Co & 2a & 0 & 0 & 0 & 0.09(17) \\
Os & 2b & 0 & 0 & 0.5 & 0.18(6)\\
O1 & 4e & -0.0679(4) & -0.0088(7) & 0.2592(4) & 0.53(6) \\
O2 & 4e & 0.2413(6) & 0.2810(7) & 0.0362(8) & 0.55(12) \\
O3 & 4e & 0.2834(6) & 0.7651(6) & 0.0365(8) & 0.50(12) \\
\hline
Site &  & M$_a$ & M$_b$ & M$_c$ & irrep \\ 
\hline
Co &  & -0.77(1) & 0 & 1.89(1) & mY$_1^+$ \\
Os &  &  -0.77(1) & 0 & 1.89(1) & mY$_2^-$\\
\hline
\end{tabular}
\label{table:1}
\end{table}

\begin{figure}
\includegraphics[width=0.45\textwidth]{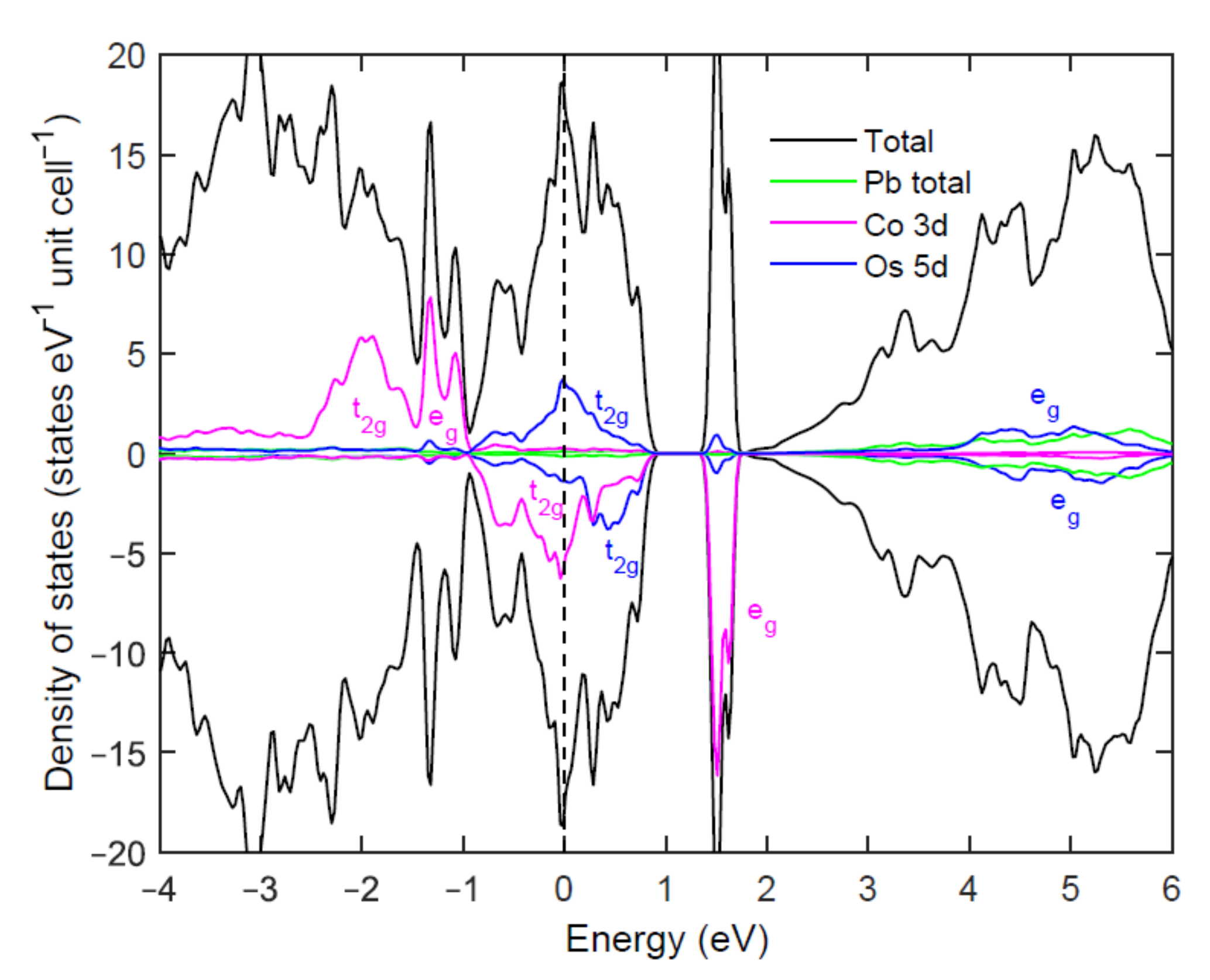}
\caption{\label{Fig4}  Total and projected spin-polarised partial density of states (DOS) for antiferromagnetic Pb$_2$CoOsO$_6$ calculated in the GGA approximation. Dotted (magenta) lines are for the Co 3$d$ orbitals and dashed (blue) lines are the Os 5$d$ orbitals. Those contributions from the crystal field split t$_{2g}$ and e$_g$ orbitals are also labeled. The Fermi energy is set at zero. Positive and negative values of the DOS represent the spin-up and spin-down parts, respectively. }
\end{figure}

\section{Electronic Structure Calculations}
The electronic structure of Pb$_2$CoOsO$_6$ was investigated by performing first-principles calculations with the OpenMX software package \cite{openMX} using the lattice parameters and the AFM configuration obtained from the NPD refinement. The choice of pseudo-potentials, basis sets and the sampling of Brillouin zone with 6$\times$6$\times$10 grid have been carefully optimized and the exchange-correlation functional within the generalized gradient approximation (GGA) \cite{Perdew1996} was used. For Co 3$d$ orbitals, the spin splitting is larger than that due to the CF, so that Co$^{2+}$ is in its high spin state with configuration as t$_{2g}^{3\uparrow}$ e$_g^{2\uparrow}$  t$_{2g}^{2\downarrow}$  e$_g^{0\downarrow}$ , where the superscript number represents the number of occupied electrons and arrows indicate the spin state. On the contrary, the extended 5$d$ orbitals of Os have a larger CF splitting than spin splitting, which consequently results in a low spin state of Os$^{6+}$ (5$d^2$) with an approximate configuration of t$_{2g}^{1\uparrow}$  e$_g^{0\uparrow}$  t$_{2g}^{1\downarrow}$  e$_g^{0\downarrow}$ . The calculated density of states (DOS) indicates substantial contributions from both Co and Os $d$-electrons at the Fermi level, both of which are split into t$_{2g}$ and e$_g$ manifolds under the approximately octahedral crystal field (CF). This large density of states at the fermi level is consistent with the robust metallicity we observe experimentally. The calculated local magnetic moments inferred from this are 2.46 $\mu_{\rm B}$ and 0.450 $\mu_{\rm B}$, on Co and Os respectively, consistent with the above approximate atomic configuration and with the range of possible experimental values indicated by NPD [see insert to Figure 3(a)]. 

\section{Muon Spin rotation and DFT+$\mu$}
The scenario of a magnetically-induced structural distortion described above requires magnetic order on both the Co and Os sublattices. As a further test of this  picture we turned to zero-field muon-spin rotation ($\mu^+$SR), which is a local probe of magnetism. Zero-field (ZF) muon spin relaxation ($\mu^+$ SR) spectra of Pb$_2$CoOsO$_6$ were measured in a $^3$He cryostat in the General Purpose Spectrometer (GPS) at the Swiss Muon Source at the Paul Scherrer Institute, Switzerland. The muon spectra (Figure~\ref{plot:asymsfit}, blue curve) below the transition can be well modelled with the equation
\begin{equation}
A=A_1\cos{(\omega_1 t)} e^{-\lambda_1t}+A_2\cos{(\omega_2 t)} e^{-\lambda_2t}+A_3 e^{-\lambda_3t}
.\label{eq:asymsfit}
\end{equation}
The first two terms of Eq.~\ref{eq:asymsfit} represent muons stopping in the sample and experiencing long-range magnetic order with precession frequencies $\omega_i=\gamma_\mu\rm{B}_i/(2\pi)$ and relaxation rates $\lambda_i$, where $\gamma_\mu=2\pi\times135.5\rm{MHzT}^{-1}$ is the muon muon gyromagnetic ratio and $\rm{B}_i$ are the local fields. The third term accounts for muons stopping outside the sample, such as in the sample holder. The results for $\lambda_i$ and $\nu_i$ are plotted in Figure~\ref{plot:asymsfit}, together with phenomenological order parameter models of the form $\nu=\nu_0[1-(T/T_{\rm N})^\alpha]^\beta$ from which a critical temperature of $T_{\rm N}=45.5(10)$~K can be extracted. Furthermore, the ratios of the amplitudes $A_i$ averaged over the measured temperatures below $T_{\rm N}$ indicate that about 35\% of the muons in our experiments experienced the smaller magnetic field, about 60\% the higher field, and about 5\% implanted outside the sample.

\begin{figure}[htb] 
\center
\includegraphics[width=0.9\columnwidth, clip, trim= 0.0mm 0.0mm 0.0mm 0.0mm]{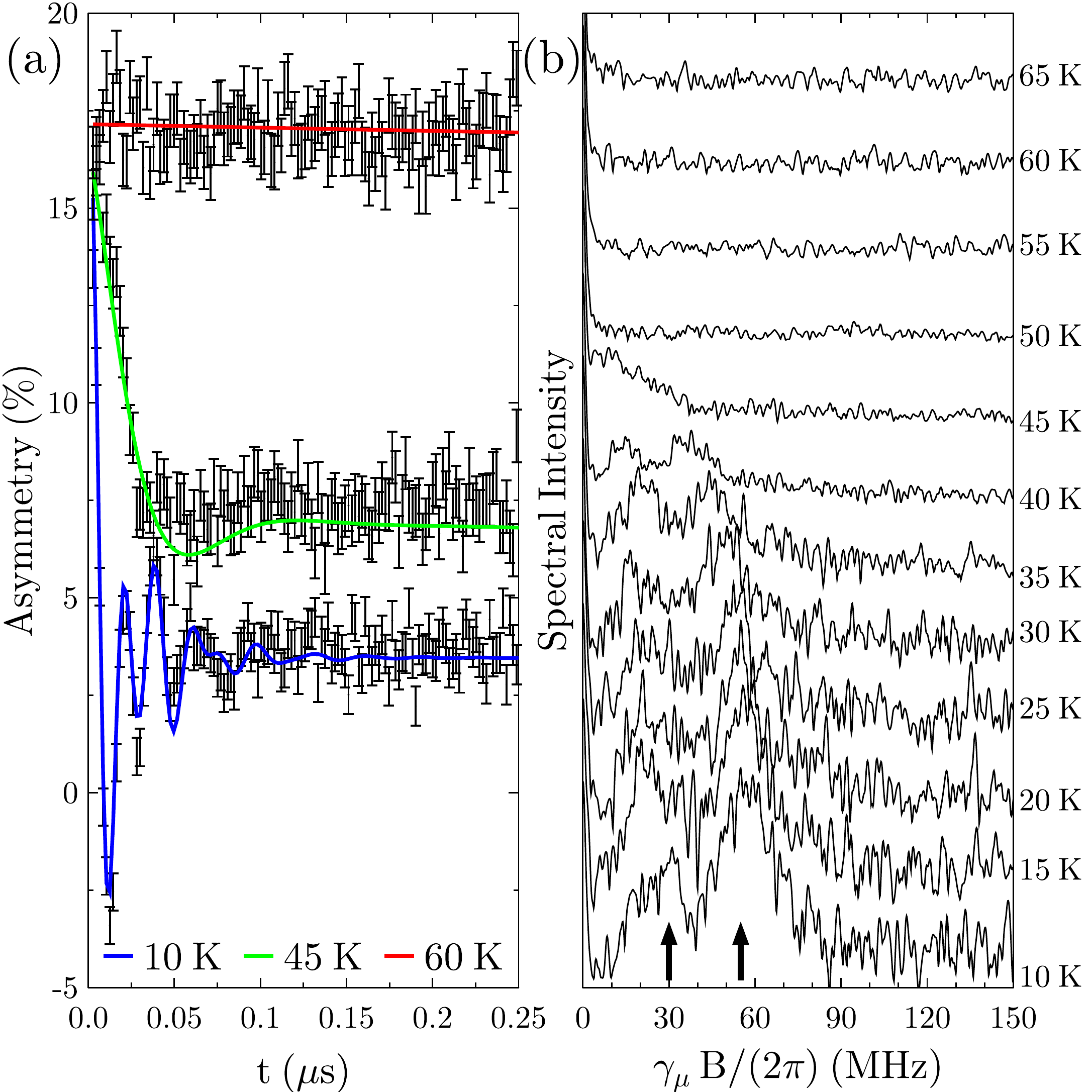}
\caption{ \label{plot:asymsandfft}
(a) ZF-$\mu^+$SR asymmetry spectra below, near and above the transition in Pb$_2$CoOsO$_6$. The blue and green lines represent fits as in Eq.~\ref{eq:asymsfit}, the red line a Lorentzian relaxation fit of the form $A\exp(-\lambda t)$. (b) Fourier transform spectra of the first 0.5~$\mu$s of the muon asymmetries (vertically shifted for readibility). The arrows indicate the position of the two frequencies at 10K
}
\end{figure}

\begin{figure}[htb] 
\center
\includegraphics[width=0.75\columnwidth, clip, trim= 0.0mm 0.0mm 0.0mm 0.0mm]{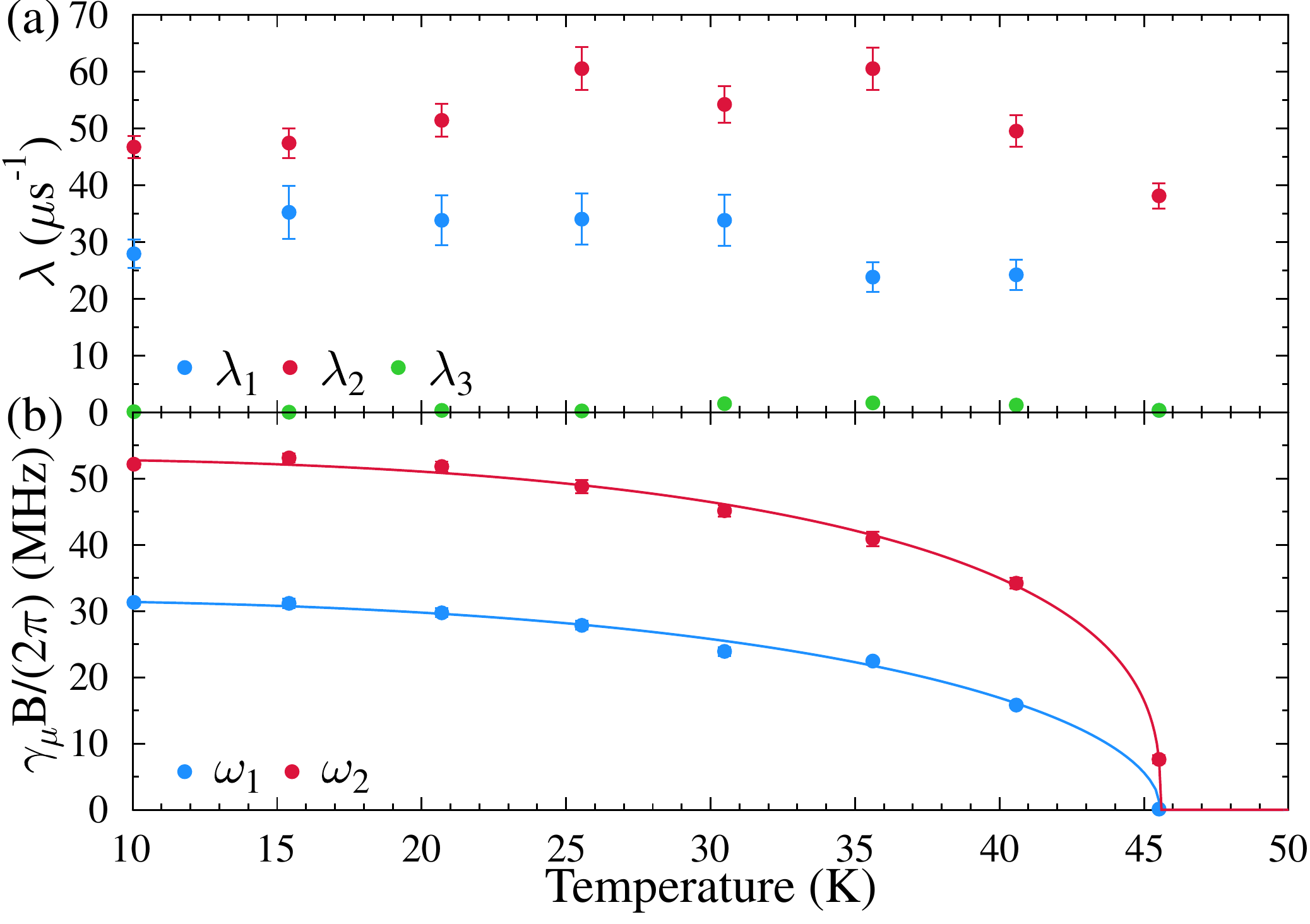}
\caption{ \label{plot:asymsfit}
Results of fitting Eq.~\ref{eq:asymsfit} to the ZF-$\mu^+$SR assymetry spectra. Solid lines in (b) represent phenomenological order parameter fits of the form $\nu=\nu_0[1-(T/T_{\rm N})^\alpha]^\beta$.
}
\end{figure}

\begin{figure}[htb] 
\center
\includegraphics[width=0.75\columnwidth, clip, trim= 0.0mm 0.0mm 0.0mm 0.0mm]{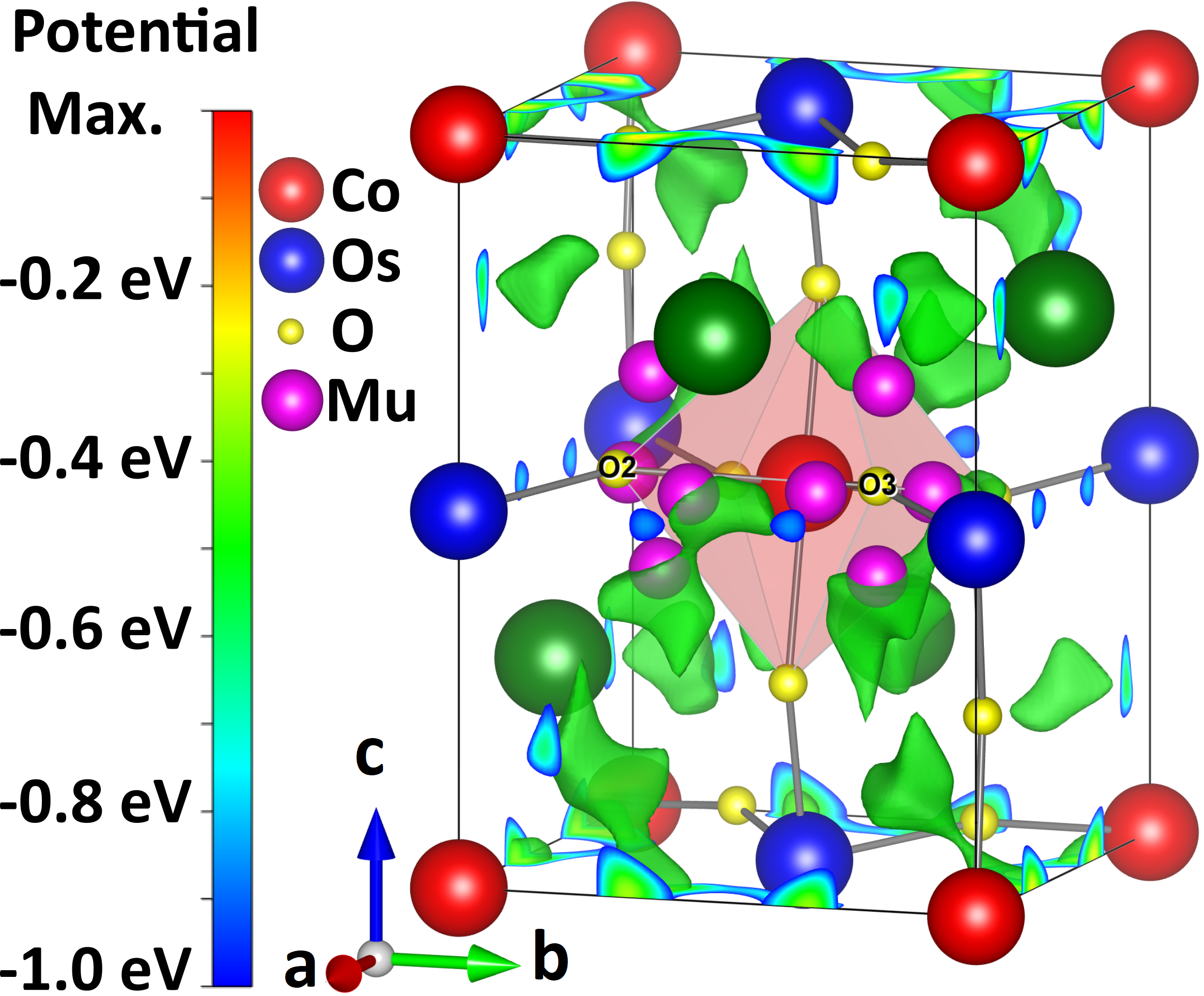}
\caption{ \label{plot:coulomb}
Electrostatic Coulomb potential of Pb$_2$CoOsO$_6$ computed with DFT. The potential is shown on the surface of the unit cell up to $1.0$\,eV below its maximum value, and a green isosurface is plotted within the unit cell at $0.5$~eV below the maximum. In addition, the eight muon site candidates (or a symmetry equivalent point) obtained from DFT relaxation calculations are also plotted in this undistorted unit cell to highlight the close symmetry relation between them. Note that the Pb ions are excluded for readibility reasons.
}
\end{figure}

In order to compare the observed muon precession frequencies with ones predicted from different magnetic structures, it is necessary to establish the potential muon stopping sites in Pb$_2$CoOsO$_6$. To this end we employed Density Functional Theory (DFT) calculations using the plane-wave program {\it Quantum Espresso}~\cite{Gianozzi2009} within the generalized gradient approximation (GGA)~\cite{Perdew1996}. We modelled the ions with ultrasoft pseudopotentials~\cite{Rappe1990} and the muon with a norm-conserving hydrogen pseudopotential. The energy cutoffs for the wavefunction and the charge density were set to $80$\,Ry and $800$\,Ry, respectively, and a $3\!\times\! 3\! \times \!3$ Monkhorst-Pack $k$-space grid~\cite{Monkhorst1976}  was used for the integration over the Brillouin zone. Within these parameters the calculations gave well converged results and reproduced the experimentally observed atomic positions and lattice parameters within a 1\% accuracy. The results shown in Figure~\ref{plot:coulomb} were visualised with the Vesta software~\cite{Momma2008}.

We use the converged electron density to map out the electrostatic Coulomb potential of Pb$_2$CoOsO$_6$ throughout its unit cell, as plotted in Figure~\ref{plot:coulomb}. The global maximum is used as the reference point, since large values of the Coulomb potential correspond to a low energy cost to add a positive charge and such regions have been found to be a reliable first-order estimate of potential muon sites~\cite{Moller2013_DFT,Foronda2015}. Addtionally, we performed relaxation calculations, which allow for distortions of the lattice due to the implanted muon, which lead to muon site candidates in good agreement with the sites predicted by the local maxima of the electrostatic potential. In fact, we find that there are eight potential muon site candidates, which are all fairly closely related in energy and symmetry due to the proximity of the crystal structure of Pb$_2$CoOsO$_6$ to a more symmetric cubic one. These muon sites, plotted in Figure~\ref{plot:coulomb} and tabulated in Table~\ref{tab:dftmu}, are also all characterised by an O--H like bond between the muon and an oxygen with a bond length of about $\SI{1.0}{\angstrom}$, which is a typical occurence in oxygen containing compounds \cite{Moller2013_DFT,Foronda2015}.

 The muon asymmetry, plotted in Figure~\ref{plot:asymsandfft}(a) for three temperatures, exhibits an oscillatory beating pattern at low temperatures indicative of long-range magnetic ordering. The Fourier transform spectra of the ZF-$\mu^+$SR asymmetries, presented in Figure~\ref{plot:asymsandfft}(b), reveals two broad peaks centered around muon precession frequencies of roughly 30~MHz and 55~MHz at the lowest measured temperatures which vanish above the transition temperature. Likely muon stopping sites in the magnetic unit cell were calculated using DFT methods, and we find 8 possible muon stopping sites with comparable energies, which could thus potentially be occupied. We label these sites Mu1-8 in order of increasing enregy cost for occupation and thus decreasing probability of occupation. 

\begin{table}[htbp]
\begin{tabular*}{\columnwidth}{@{}l  @{\extracolsep{\fill}}*3c @{\extracolsep{\fill}}*2c l@{}}
\hline
  Site & \multicolumn{3}{c}{Relaxed position} & $\Delta$E/u.c.  & O--H \\
  Label & x & y & z  & (meV) & with\\
\hline
Mu1 & 0.786 & 0.225 & 0.838 & 0 & O3 \\
Mu2 & 0.264 & 0.720 & 0.334 & 4.5 & O2 \\
Mu3 & 0.295 & 0.878 & 0.026 & 6.5 & O3 \\
Mu4 & 0.878 & 0.699 & 0.983 & 24.5 & O2 \\
Mu5 & 0.880	& 0.288 & 0.988 & 24.8 & O3 \\
Mu6 & 0.305 & 0.121 & 0.015 & 40.5 & O2 \\
Mu7 & 0.710 & 0.291 & 0.403 & 46.6  & O2 \\
Mu8 & 0.710 & 0.711 & 0.407 & 46.8 & O3 \\

\hline 
\end{tabular*}
\caption{ \label{tab:dftmu}
Summary of DFT+$\mu$ calculations. From left to right: site label, fractional coordinates of muon site candidates from relaxation calculation, energy difference per unit cell for different muon site configurations, label of Oxygen with which the muon forms an O-H like $\SI{1.0}{\angstrom}$ bond.
}
\end{table}

\begin{figure}[htb] 
\center
\includegraphics[width=\columnwidth, clip, trim= 0.0mm 0.0mm 0.0mm 0.0mm]{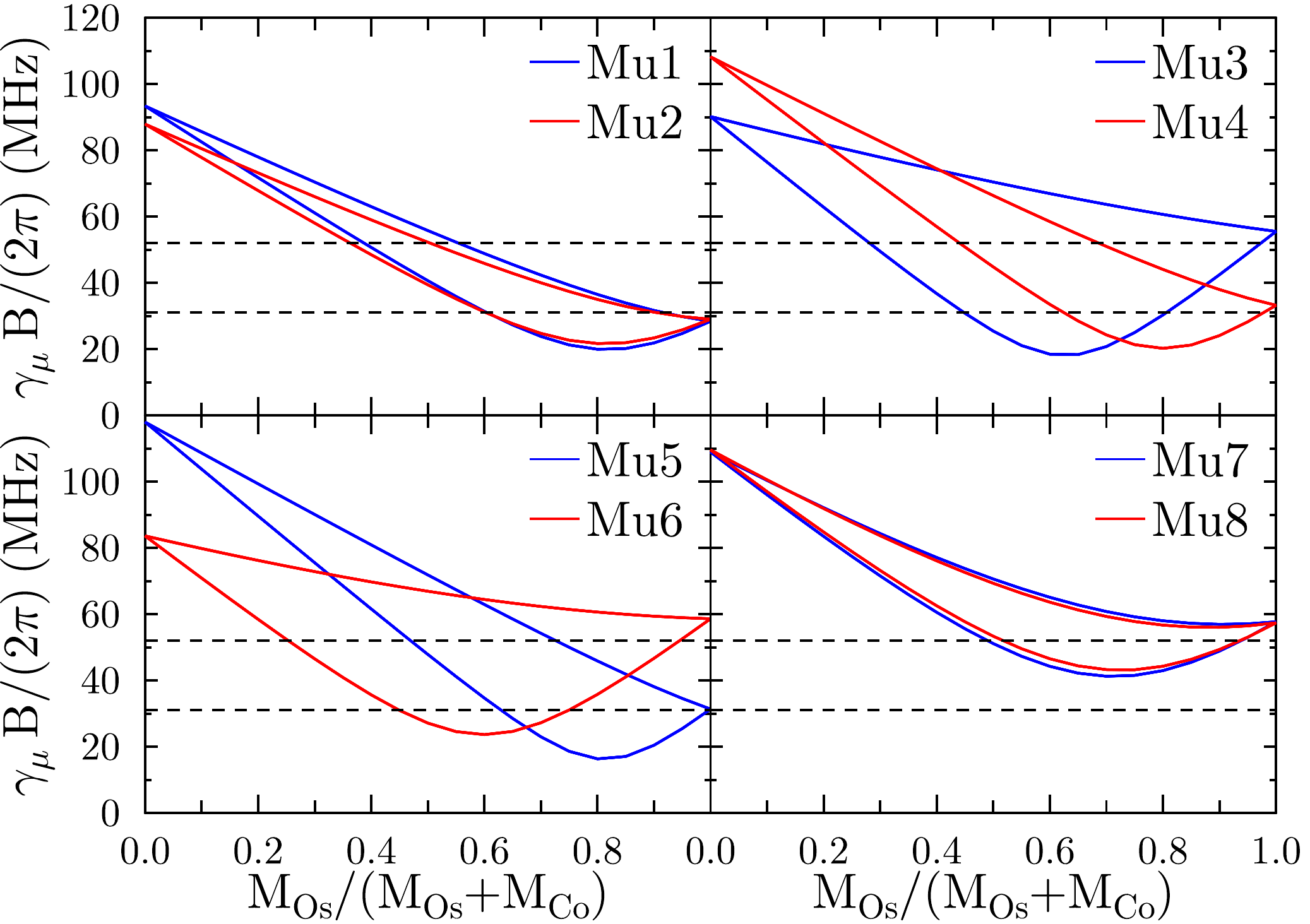}
\caption{ \label{plot:dipolefields}
Muon precession frequencies computed from the dipolar fields expected at the DFT-calculated muon sites. The magnetic structure is taken to be that shown in Figure~3~(b) of the main text, and the total moment size that in Table~1. The fraction of the Os moment to the total moment was varied in line with the inset of Figure~3~(a). The muon site labels correspond to those in Table~\ref{tab:dftmu}. Note that the frequencies at all 16 symmetry equivalent positions for each muon site are plotted, which results in pairs of eight very similary, and visually indistinguishable curves in each case.
} 
\end{figure}

With the potential muon spotting sites from Table~\ref{tab:dftmu}, we can calculate the local magnetic fields at each of these sites. Because our neutron experiments revealed an antiferromagnetic ordering the Lorentz and demagnetising fields are zero and since the muon sites are far from the Os and Co ions we also expect hyperfine field contributions to be negligible, such that we only have to focus on the dipolar fields
\begin{equation}
\boldsymbol{{B}}_\mathrm{local}\approx\boldsymbol{{B}}_\mathrm{dip}
=\sum_i \frac{\mu_o}{4\pi |\Delta\boldsymbol{r}_i|^3}\left[\frac{3(\boldsymbol{\mu}_i\cdot\Delta\boldsymbol{r}_i)\Delta\boldsymbol{r}_i}{|\Delta\boldsymbol{r}_i|^2}-\boldsymbol{\mu}_i\right]
\nonumber.\end{equation}
Here, the $\boldsymbol{r}_i$ correspond to the relative positions of the magnetic moments $\boldsymbol{\mu}_i$ with respect to the muon. Using the total moment size, as listed in Table~1, we computed the local magnetic fields, assuming a fraction $x$ of the total moment lies on the Os sites, while a fraction $1-x$ of the total moment lies on the Co sites, for each of the muon site candidates and their 16 symmetry equivalent sites in the magnetic unit cell. The resulting muon precession frequencies are plotted in Figure~\ref{plot:dipolefields}. Note that each line in Figure~\ref{plot:dipolefields} actually represents eight nearly identical lines, such that the 16 positions symmetry equivalent to the Mu1 site actually only lead to two distinct experimentally observable frequencies (unless the Os moment fraction $x$ is very close to 0 or 1). We can further note that the local fields at some of the muon stopping sites are almost identical, which is owed to the proximity to a cubic crystallographic symmetry (see Figure~\ref{plot:coulomb}). Based on the muon precession frequencies shown in Figure~\ref{plot:dipolefields} we can identify two scenarios which can plausibly explain our experimentally observed frequencies. Either, only the two energetically most favourable sites (Mu1 and Mu2) are significantly occupied and that $x \approx 0.5$. Alternatively, all eight of the identified muon site candidates might be occupied and $x \approx 1$. Both scenarios would lead to only two experimentally distinguishable frequencies, with a ratio of the oscillation amplitudes of these two frequencies expected in the region of 1:1, very roughly in line with that observed experimentally. 

\section{Discussion}
Of the two realistic possibilities outlined above, the scenario with $x \approx 1$ (i.e. the full moment is on the osmium site) is very unlikely. First, this scenario would require that the cobalt fails to order, contrary to other known double perovskites of mixed 3$d$-5$d$ cations, in which the 3$d$ cation is either the only element that orders or the first to order by a significant margin. Second, the Os$^{6+}$ in Pb$_2$CoOsO$_6$ would have to carry a moment (approx 4$\mu_{\rm B}$) substantially greater than that which has been observed in any osmium containing compound to date; and indeed greater than should be possible even for an ideal $J=2$ case (giving 3.6$\mu_{\rm B}$), which anyway does not seem to be favoured for Os due to covalency effects that are substantial compared to 3d and even 4d elements\cite{gangopadhyay,taylor, BCOO}. The likelihood of such a large moment on Os is further diminished by recent results on other 5$d^2$ osmium double-perovskites Ba$_2$CaOsO$_6$ (ordered moment $<$ 0.2$\mu_{\rm B}$ \cite{BCOO}), Sr$_2$CoOsO$_6$ (ordered moment 1.8$\mu_{\rm B}$ \cite{morrow_jacs, magnetoelasticosmate}), Ca$_3$OsO$_6$ (moment estimated from curie--weiss fit $<$ 2$\mu_{\rm B}$ \cite{COO}), and both Sr$_2$MgOsO$_6$ and Ca$_2$MgOsO$_6$ (ordered moment $\approx$ 1.8$\mu_{\rm B}$\cite{CMOO}). The weight of evidence overwhelmingly suggests that both the Os and Co ions carry an ordered moment in the magnetically ordered phase. 

\section{Conclusion}
We conclude that the antiferromagnetic phase transition in Pb$_2$CoOsO$_6$, involving the simultaneous magnetic order on both the Co and Os sublattices, removes the centre of crystal symmetry and through magnetoelastic coupling relaxes the structure into the polar multiferroic Shubnikov group P$_a$c. In Pb$_2$CoOsO$_6$, the spin-driven nature of the structural transition is reminiscent of a type-II hybrid improper multiferroic transition, in which magnetic order removes inversion symmetry via some higher order invariant in the free energy \cite{hybridimproper1,hybridimproper2}. A key difference between the present case and all known multiferroics, however, is that the transition in Pb$_2$CoOsO$_6$ occurs entirely in the metallic state. Oxide NCSMs are still very rare, and the materials known up to now exist because the polar displacements are almost entirely decoupled from the conduction electrons. The magnetic mechanism for acentricity in Pb$_2$CoOsO$_6$ does not suffer from this constraint and could provide a new route for the discovery of other oxide NCSMs and their associated novel phenomena. Thus, the phase transition we have illustrated in Pb$_2$CoOsO$_6$ could provide a new guiding principle for the future discovery of oxide NCSMs.

\section{Acknowledgments}

Two authors (AJP and HLF) contributed equally to this work. This research was supported in part by the World Premier International Research Center from MEXT; the Grants-in-Aid for Scientific Research (25289233) from JSPS; the Funding Program for World-Leading Innovative R\&D on Science and Technology from JSPS; and United Kingdom Engineering and Physical Sciences Research Council (EPSRC). H.M.W. acknowledges the supports from NSF of China and the 973 program of China (No. 2011CBA00108 and 2013CB921700). YFG acknowledges the support of Shanghai Pujiang Program, grant No. 17PJ1406200. We performed the SXRD measurements with the approval of the NIMS beamline station (Proposal No. 2013B4503).  This work was supported by EPSRC (UK) grants EP/N034872 and EP/N023803. The authors acknowledge the use of the University of Oxford Advanced Research Computing (ARC) facility (http://dx.doi.org/10.5281/zenodo.22558).

\bibliography{PCOO}

\end{document}


\title[Supplementary Material for:  Magnetically driven loss of centrosymmetry in metallic Pb$_2$CoOsO$_6$]{Supplementary Material for: Magnetically driven loss of centrosymmetry in metallic Pb$_2$CoOsO$_6$}

\author{A. J. Princep}
\email[]{princep@physics.ox.ac.uk}
\affiliation{Department of Physics, University of Oxford, Clarendon Laboratory, Parks Road, Oxford, OX1 3PU, United Kingdom}
\affiliation{ISIS Facility, Rutherford Appleton Laboratory, Chilton, Didcot, OX11 0QX, United Kingdom}
\author{H. L. Feng}
\email[]{hai.feng@iphy.ac.cn}
\affiliation{National Institute for Materials Science, 1-1 Namiki, Tsukuba, Ibaraki 305-0044, Japan}
\affiliation{Beijing National Laboratory for Condensed Matter Physics, Institute of Physics, Chinese Academy of Sciences, Beijing 100190, China}
\author{Y. F. Guo}
\email[]{guoyf@shanghaitech.edu.cn}
\affiliation{School of Physical Science and Technology, ShanghaiTech University, Shanghai 201210, China}
\author{F. Lang}
\affiliation{Department of Physics, University of Oxford, Clarendon Laboratory, Parks Road, Oxford, OX1 3PU, United Kingdom}
\author{H. M. Weng}
\affiliation{Beijing National Laboratory for Condensed Matter Physics, Institute of Physics, Chinese Academy of Sciences, Beijing 100190, China}
\affiliation{Collaborative Innovation Center of Quantum Matter, Beijing, China }
\author{P. Manuel}
\author{D. Khalyavin}
\affiliation{ISIS Facility, Rutherford Appleton Laboratory, Chilton, Didcot, OX11 0QX, United Kingdom}
\author{A. Senyshyn}
\affiliation{FRM-II, Technische Universität München, Garching 85747, Germany}
\author{M. C. Rahn}
\affiliation{Department of Physics, University of Oxford, Clarendon Laboratory, Parks Road, Oxford, OX1 3PU, United Kingdom}
\affiliation{Institute for Solid State and Materials Physics, Technical University of Dresden, 01062 Dresden,Germany}
\author{Y. H. Yuan}
\author{Y. Matsushita}
\affiliation{National Institute for Materials Science, 1-1 Namiki, Tsukuba, Ibaraki 305-0044, Japan}
\author{S. J. Blundell}
\affiliation{Department of Physics, University of Oxford, Clarendon Laboratory, Parks Road, Oxford, OX1 3PU, United Kingdom}
\author{K. Yamaura}
\email[]{Yamaura.kazunari@nims.go.jp}
\affiliation{National Institute for Materials Science, 1-1 Namiki, Tsukuba, Ibaraki 305-0044, Japan}
\affiliation{Graduate School of Chemical Sciences and Engineering, Hokkaido University, Sapporo, Hokkaido 060-0810, Japan}
\author{A. T. Boothroyd}
\email[]{a.boothroyd@physics.ox.ac.uk}
\affiliation{Department of Physics, University of Oxford, Clarendon Laboratory, Parks Road, Oxford, OX1 3PU, United Kingdom}

\date{\today}



\maketitle

\section{Preparation and characterization}

Both poly- and single-crystalline Pb$_2$CoOsO$_6$ were prepared by solid-state reaction in a belt-type high-pressure apparatus \cite{synthesis1,synthesis2}. A mixture of PbO$_2$ (3N, High Purity Chemicals. Co., Ltd., Japan), Os (99.95\%, Heraeus Materials Technology, Germany), Co (4N, Alfa Aesar), and KClO$_4$ ($>$ 99.5\%, Kishida Chemical Co., Ltd., Japan) in the molar ratio PbO$_2$/Os/Co/KClO$_4$ = 2:1:1:0.5 was placed in a platinum capsule (6.9 mm in diameter and $\sim$ 5 mm in height). The capsule was loaded into the high-pressure apparatus and treated in a pressure of 6 GPa at 1500 $^\circ$C for 1 hour (polycrystals) or at 1600 $^\circ$C for 2 hours (single crystals). After heating, the capsule was quenched to room temperature within a few seconds before releasing the pressure. The products were then rinsed in water for 5 min for several times to remove residual ingredients. Phase purity of the powder sample was examined by the synchrotron x-ray diffraction (SXRD) at room temperature using the large Debye-Scherrer camera at the BL15XU beamline in SPring-8, Japan \cite{spring8}. The data were collected between 2$^\circ$ and 60$^\circ$ with a 0.003$^\circ$ step in 2$\Theta$ and an incident wavelength of $\lambda$ = 0.65297 $\AA$. Fine powder of the sample was packed into a 0.1 mm Lindenmann glass capillary which was rotated during measurement. The obtained SXRD patterns were analyzed by the Rietveld method using the programs RIETAN-2000 and VESTA \cite{RIETAN,VESTA}. The refinement results  are presented in Fig. S1 and Table S1.  Laboratory X-ray diffraction was performed using a Mo-source Oxford Diffraction Supernova diffractometer on a single crystal of Pb$_2$CoOsO$_6$ of approximate size 820$\times$310$\times$110$\mu$m$^3$ (Figure S2a). More than 96\% of the detected peaks were successfully indexed by a single monoclinic domain with the space group P2$_1$/$n$. The diffraction patterns along the [0 k l], [h 0 l], and [h k 0] directions are shown in Figs. S2b-2c. The small thermal parameters of Osmium and Cobalt are most likely the result of the refinement  parameters which describe absorption being strongly coupled to the thermal displacement parameters. The osmium-cobalt site disorder was refined in the high-temperature diffraction data. We found no evidence for site disorder to $< 1 \%$, which was the limit of sensitivity of the measurement. 

Specific heat was measured in a Quantum Design PPMS. The peak in the specific heat shows a slight downwards shift ($<$ 1 K) in an applied magnetic field of 90 kOe. To estimate the magnetic entropy change (S$_{\rm  Mag}$), we subtracted the lattice contribution (C$_{\rm lat}$) from C$_{\rm p}$ via a polynominal fit. We estimate S$_{\rm Mag}$ to be approximately 3.65 J mol$^{-1}$ K$^{-1}$. This value is less than 20$\%$ of the expected value of 20.7 J mol$^{-1}$ K$^{-1}$ for spin-only Os and Co. The low-temperature part of Cp was analyzed separately using the Debye model with an electronic specific heat term (Fig. S2b), i.e. C$_{\rm p}$/T = $\gamma$ + $\beta$T$^2$, where $\gamma$ is the Sommerfeld constant. The analysis yielded a $\gamma$ value of 33.6(6) mJ mol$^{-1}$ K$^{-2}$, which is consistent with a substantial density of states at the Fermi level. In a magnetic field of 90 kOe, a similar analysis revealed negligible change of $\gamma$. 

To parameterize C$_{ \rm p}$, the curve was analyzed by a linear combination of the Debye and the Einstein models (see the solid line in Fig. 2c): 
\begin{equation}
\begin{split}
\frac{C_p(T)}{n_D} = 9N_Ak_B ( \frac{T}{T_D} )^3 \int_0^{T_D/T} \frac{x^4e^x}{( e^x -1 )^2} \\+  3N_Ak_B ( \frac{T_E}{T} )^2  \frac{e^{T_E/T}}{( e^{T_E/T} -1 )^2} ,\label{eq:1}
\end{split}
\end{equation}
where N$_A$ is the Avogadro’s constant, k$_B$ is the Boltzmann’s constant, and T$_E$ and T$_D$ are the Einstein and Debye temperatures, respectively. The scale factors n$_D$ and n$_E$ correspond to the number of vibrating modes per the formula unit in the Debye and the Einstein models, respectively. The fit yielded T$_D$ of 610(7) K, T$_E$ of 90(12) K, n$_D$ of 6.997(4), and n$_E$ of 3.535(5). The non-trivial Einstein term likely suggests anharmonic lattice dynamics in Pb$_2$CoOsO$_6$.

\begin{table}
\caption{Structural parameters of Pb$_2$CoOsO$_6$ from synchrotron x-ray diffraction at room temperature. Space group P2$_1$/$n$ ( \#14, setting choice 2), Z = 2, $\lambda$ = 0.65297 $\AA$; Full occupancy factors for all sites; a = 5.6116(41) $\AA$, b = 5.65743(7) $\AA$, c = 7.9049(21) $\AA$, and $\beta$ = 89.990(33)$^\circ$; Final R values are: R$_p$ = 1.095\% and R$_{wp}$ = 1.778\%.}
\begin{tabular}{ c c c c c c } 
 \hline
Site & Wyckoff & x & y & z & B$_{iso}$ ($\AA^2$) \\ 
\hline
Pb & 4e & 00.0032(4) & 0.5057(5) & 0.25071(5) & 1.14252 \\
Co & 2a & 0 & 0 & 0 & 0.328610 \\
Os & 2b & 0 & 0 & 0.5 & 0.477756 \\
O1 & 4e & -0.0606(5)&	-0.0019(4)	&0.2587(11)	&0.99996 \\
O2 & 4e & 0.2399(5)&	0.2817(3)	&0.0297(7)&	0.99996 \\
O3 & 4e & 0.2797(4) &	0.7610(3)&	0.0327(5)	&0.99996 \\
\hline
\end{tabular}
\label{table:2}
\end{table}

\begin{table}
\caption{Selected interatomic distances, angles, and Bond Valence Sums (BVS) of Pb$_2$CoOsO$_6$ at room temperature. BVS =  $\sum_{i=1}^N v_i$ $v_i = \exp{(R_0 - l_i)/B}$, N is the coordination number, B = 0.37, R$_0$(Co$^{2+}$) = 1.692, and R$_0$(Pb$^{2+}$) = 2.1124, and R$_0$(Os$^{6+}$) = 1.925.}
\begin{tabular}{ c c c c } 
\hline
Bond & Bond distance ($\AA$) & Bond & Bond angle ($^\circ$) \\
\hline
Co1-O1 & 1.9027 (6) $\times$ 2	& Os-O1-Co & 172.2(8) \\
Co1-O2 & 2.1084(3) $\times$ 2 & Os-O2-Co & 168.1(7)\\
Co1-O3 & 2.2409(2) $\times$ 2 & Os-O3-Co & 145.4(5)\\
BVS & 2.19251 &  & \\  
Os2-O1 & 1.8258(13) $\times$ 2 &  & \\
Os2-O2 & 1.8857(11) $\times$ 2 &  & \\
Os2-O3 & 2.1119(6)  $\times$ 2 & & \\
BVS & 6.04595 &  &  \\
\hline
\end{tabular}
\label{table:3}
\end{table}

\begin{figure}
\includegraphics[width=0.75\textwidth]{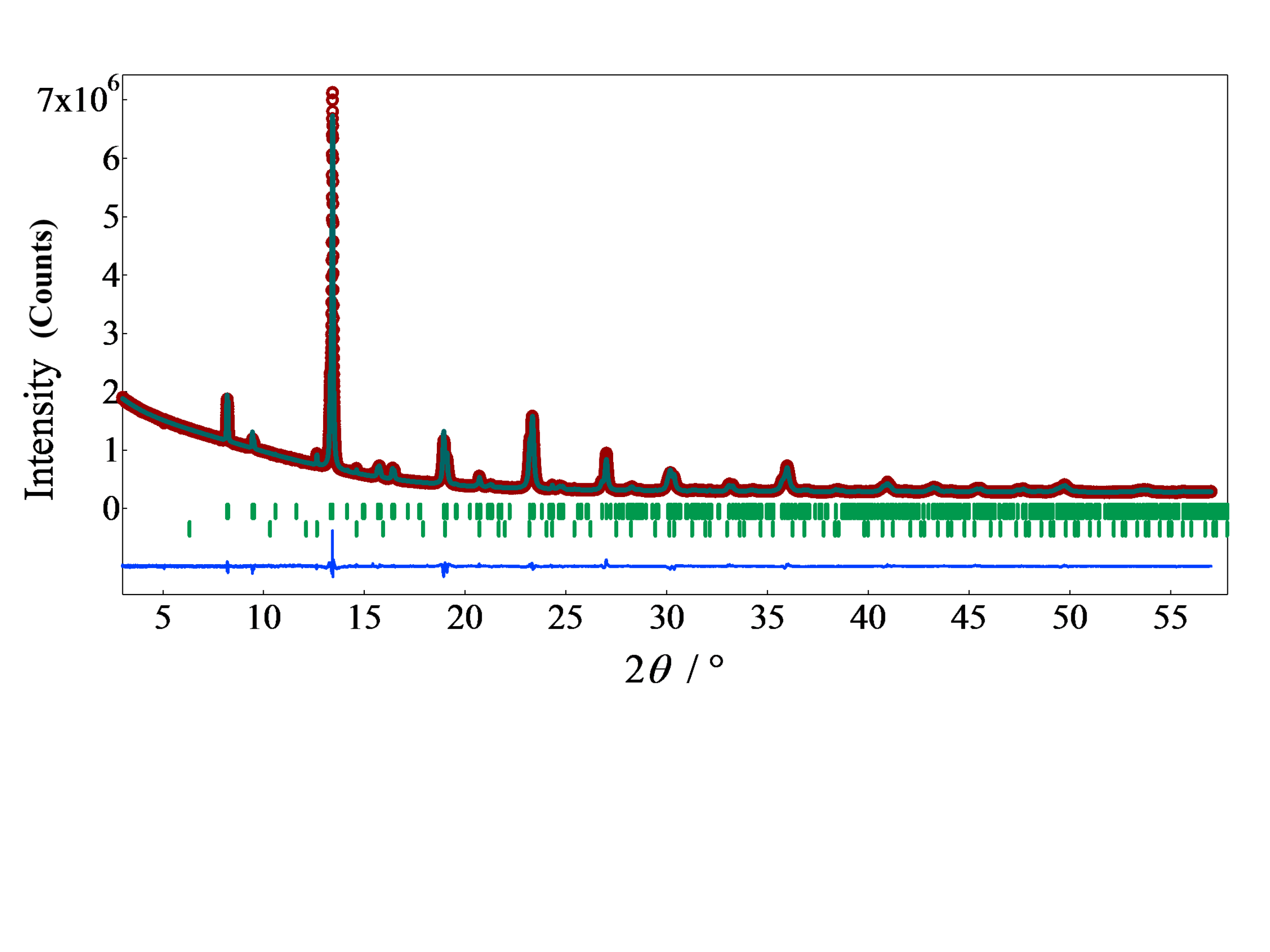}
\\[5pt]
\caption{\label{fig:S1} Fig.~S1.  Rietveld refined synchrotron XRD profiles of Pb$_2$CoOsO$_6$ at room temperature. The lower ticks show the unknown impurity.}
\end{figure}

\begin{figure}
\includegraphics[width=0.75\textwidth]{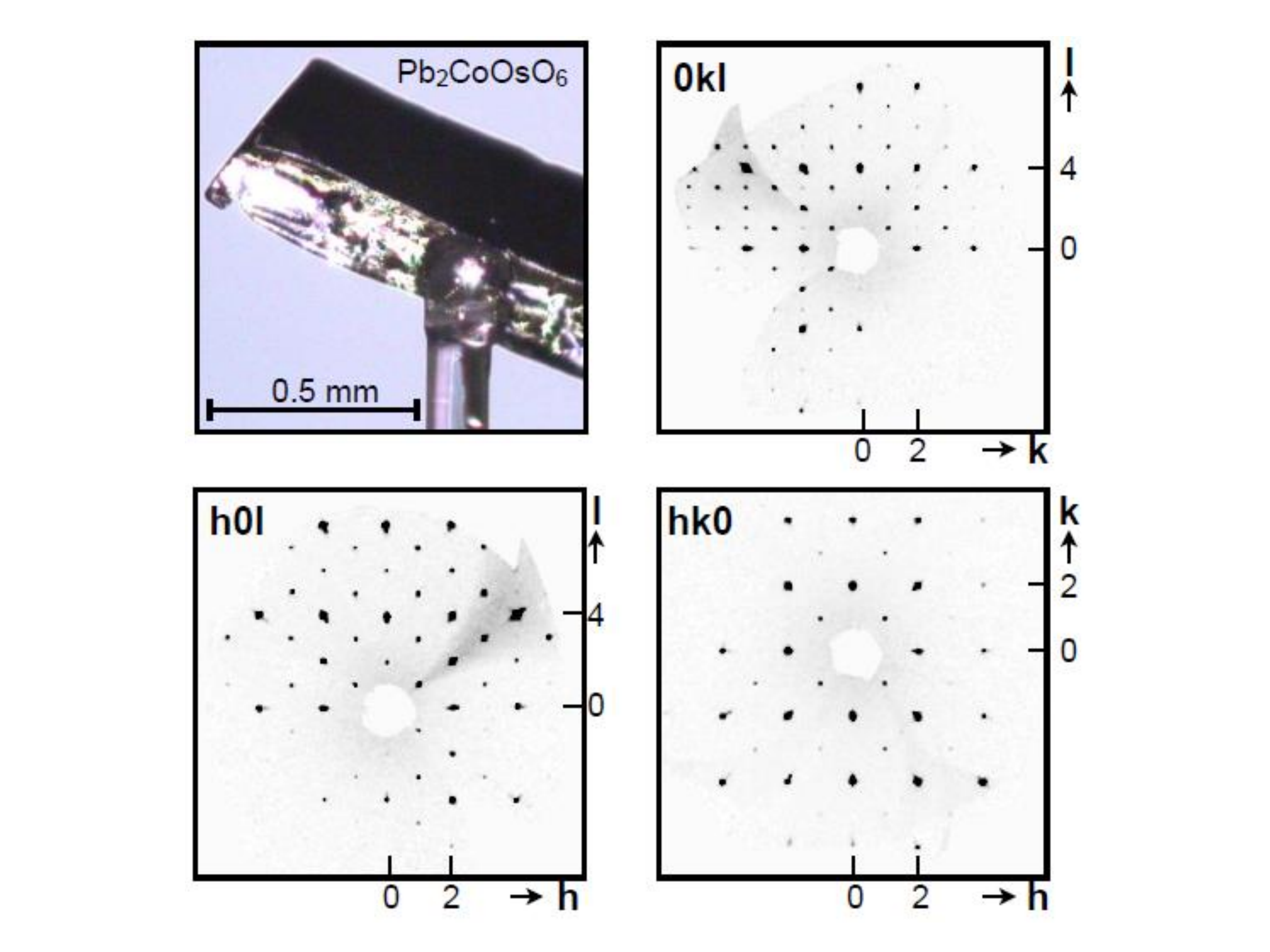}
\\[5pt]
\caption{\label{fig:S2} Fig.~S2.(a) picture of crystal used for X-ray diffraction measurement and diffraction patterns at room temperature along the (b)[0 k l], (c) [h 0 l], and (d) [h k 0] directions. The correct unit cell found is monoclinic with space group of P2$_1$/$n$ and lattice parameters being consistent with those refined from synchrotron XRD. }
\end{figure}

\begin{figure}
\includegraphics[width=0.75\textwidth]{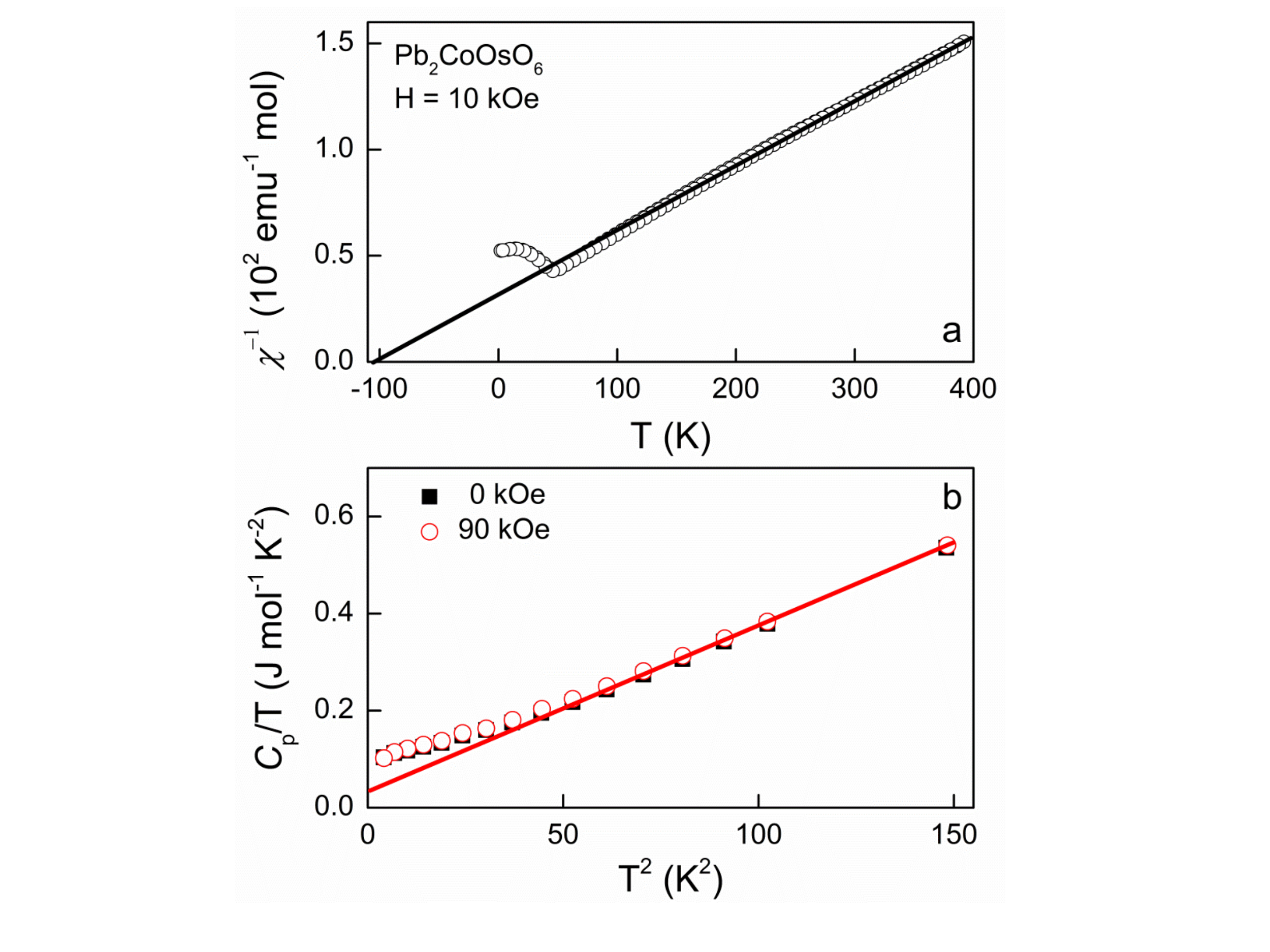}
\\[5pt]
\caption{\label{fig:S3} Fig.~S3. (a) Temperature dependence of reciprocal magnetic susceptibility ($\chi^{-1}$ vs. T) measured at H = 10 kOe. The solid line represents the result of Curie-Weiss plot ($>$ 200 K); (b) Specific heat in a form of C$_{p}$/T vs. T$^{2}$ at H = 0 and 90 kOe. The red solid line represents the plot result by using the Debye model.} 
\end{figure}

\vspace{0.6cm}

\section{Magnetic and transport properties measurements}
DC magnetic susceptibility ($\chi$) of several crystals (total mass 21.0 mg) was measured in a Quantum Design magnetic properties measurement system (MPMS) between 2 K and 400 K in an applied magnetic field (H) of 10 kOe. The crystals were loosely gathered in a sample holder and cooled down to 2 K. The magnetic field was then applied to the crystals and the temperature was slowly raised up to 400 K (Zero-Field Cooling, ZFC), followed by cooling down to 2 K again in the field (Field Cooling, FC). The isothermal magnetization was measured in the instrument at 5 K between -50 kOe and 50 kOe, which shows linear behavior.  A physical properties measurement system (PPMS) from Quantum Design was used to measure the electrical resistivity ($\rho$) of a selected crystal at temperatures between 2 K and 300 K upon cooling by a four-terminal method with an ac-gauge current of 10 mA at a frequency of 110 Hz.  Silver epoxy was used to fix platinum wires on the crystal.  Specific heat (Cp) of an amount of crystals (3.6 mg) was measured in the same apparatus between 2 K and 300 K at H = 0 and 70 kOe.

Neutron powder diffraction measurements were carried out on a 3.5 g powder sample at various temperatures on the SPODI high-resolution diffractometer \cite{SPODI} at the FRM-II facility of the Technische Universität München, Germany, and also on the WISH time-of-flight diffractometer \cite{WISH} at the ISIS Facility of the Rutherford Appleton Laboratory, UK. The Rietveld refinement of the crystal and magnetic structures was performed using FullProf \cite{fullprof}. There were only very weak impurity peaks arising from unknown impurities in the patterns (quantified as ~ 0.9 wt\%).

\bibliography{PCOO_supp}